\documentclass[12pt]{article}

\usepackage{amsmath}
\usepackage[dvips]{graphicx}
\usepackage[unicode=true]{hyperref}

\newcommand{\N}{N\raise.7ex\hbox{\underline{$\circ $}}$\;$}
\begin{document}

\title{\textbf{Zitterbewegung as purely classical
phenomenon}}
\author{A.N. Tarakanov
\thanks{E-mail: tarak-ph@mail.ru }\\
{\small Institute of Informational Technologies,}\\ {\small
Belarusian State University of Informatics and Radioelectronics}\\
{\small Kozlov str. 28, 220037, Minsk, Belarus}}
\date{}
\maketitle

\begin{abstract}
Nonrelativistic formalism is developed, which allows describing
systems with internal degrees of freedom in the scalar potential
field $U$, which is a function both on relative coordinates and
time, and on relative speed and accelerations. The equation for
energy, which is an integral of motion when $U$ satisfies to
certain differential condition, is derived for the general case.
For a free mass point all solutions of the equations of motion in
the center-of-inertia reference frame, moving with constant
velocity, are found. As a result, the center of mass follows along
helical and more complicated trajectories round a direction of
motion of the center of inertia. This motion can be interpreted as
trembling movement (Zitterbewegung). On this basis a conclusion is
done that Zitterbewegung has purely classical origin, arising even
in a nonrelativistic case if internal degrees of freedom are taken
into account. The general equation of motion for a spin which can
be interpreted from positions of a classical mechanics is written
down. Application of the obtained results to the electron leads to
new conception of electric charge, sign of which corresponds to a
sign of spin polarization.
\end{abstract}

\vskip 5mm


\section{ Introduction}

Experimental data on high energy physics say that elementary
particles are complex systems with internal degrees of freedom.
Fundamental particle which in certain relation can pretend to
simplicity, on the one hand, and underlie all known interactions,
on the other hand, is the electron. Nonexistence of the electron,
being a source of electromagnetic field, in the Maxwell theory and
its generalizations is paradoxical. It may be explained by that
the Maxwell theory describes macroscopical averaged fields created
by systems consisting of large quantity of moving charges.
Therefore the electron theory, whose foundations were laid by
Lorentz, is necessary for the description of the majority of the
electromagnetic phenomena in material systems.

The electron theory has been developing in classical trend after
discovery of the electron by J.J.Thomson in cathode rays. The
theory by Abraham~\cite{Abr}, based on the Maxwell-Lorentz
electrodynamics, became the first theory considering internal
structure of the electron. During two decades preceding experience
by Stern-Gerlach and the Kronig-Uhlenbeck-Goudsmit hypothesis
about electron spin, there were appeared a few alternative
theories and guesses about electron structure. It should be
mentioned here works by Compton~\cite{Com1}-~\cite{Com4},
influenced by both the paper by Parson~\cite{Par} and stereoscopic
photos of tracks of the $\beta$-rays, made by Wilson, some of
which had almost ideal form of a helix. Compton has come to a
conclusion that the model of the electron, "spinning like a tiny
gyroscope", can eliminate difficulties in an explanation of
curvature of these tracks, as well as of Richardson-Barnett effect
and diffraction of X-rays by magnetic crystals. Nevertheless, the
Parson-Compton theory continuing early idea of vortex atoms, going
back to Greek atomists, Kepler, Descartes, Leibnitz, Svedenborg,
Boskovich, Ampere, Kelvin, and theories of some other researchers
had rather natural philosophic and empirical nature, than they
gave any mathematical instrument for future theory. The first
mathematical realization of the Parson-Compton theory was the work
by Frenkel~\cite{Fre}, where the electron was considered as a
point with six-vector of magnetic moment, what has allowed to
explain the anomalous Zeeman effect.

After creation of quantum mechanics, especially after works by
Pauli and Dirac, the electron theory began to develop chiefly in
quantum direction. In 1930 Schr\"{o}dinger has shown that the
trembling motion (Zitterbewegung) of the electron takes place in
the Dirac theory where eigenvalues of any component of the
velocity operator are equal to $\pm c$. Microscopic trembling
motion with the velocity of light of imaginary center of the cloud
of charge, whose amplitude is about half Compton wave length, is
imposed on translational macroscopic motion of the center of mass
of the electron. As Shr\"{o}dinger talks, "exclusively entangled
relations, which are present according to the Dirac equation
already at free movement of a mass point, seem to me worthy of
enunciating though I cannot present some completed result of this
research"~\cite{Sch}. In 1952 Huang has shown that Zitterbewegung
of a free Dirac electron may be looked upon as a circular motion
about the direction of the electron spin that in turn may be
interpreted as the "orbital angular momentum" of this motion. As a
result the electric current produced by Zitterbewegung is seen to
give rise to the intrinsic magnetic moment of the electron
~\cite{Hua}.

However, as early as in 1937 Mathisson has written down general
relativistic equations of motion for systems possessed multipole
momenta~\cite{Mat1}. He has shewed also, that their application to
a particle with spin treated as a dipole gives rise to equations
of motion, describing Zitterbewegung~\cite{Mat2}. Mathisson has
considered the motion of a free uncharged particle with spin and
free electron taking account of the reaction of radiation. Here he
has assumed spin to be a constant (pseudo-)vector, what was rather
stringent assumption, for spin direction in general can change.
Nevertheless, having associated trembling motion of the electron
with the de Broglie's wave, he has obtained the well-known value
of a spin $s = \hbar / 2$.

Following basic articles by Schr\"{o}dinger and Mathisson a lot of
works were appeared, developing both quantum and classical
electron theory and establishing connection between them.
Sufficiently full list of references one may find in the
books~\cite{Cor},~\cite{Riv}. Despite a considerable quantity of
researches devoted to the theory of the electron, there are many
unclear matters associated both with radiation of the electron and
with dependence of its trajectory from its spin (see,
e.g.,~\cite{Pom}).

An interest in classical theory of the electron has especially
increased at the recent years, and this work contains some
arguments in favour of a classical origin of Zitterbewegung.
Relating of the electron spin with its proper rotation allows
considering the electron as non-inertial object which can be
described as a mass point with internal degrees of freedom
~\cite{Tar1},~\cite{Tar2}. The formalism of such description based
on generalization of the second Newton's law is considered in \S
2. Here the equations of motion of the point interacting with an
external field described by the potential function depending on
relative variables are obtained. Consequence of the equation of
motion is the equation of balance of energy which is integral of
motion only if a certain condition is fulfilled. Non-conservation
of the energy, in general, seems to be caused by the fact that
mass point in question is non-inertial system. It is supposed that
internal degrees of freedom are described by the pseudo-vectors
$\mathbf{S}$ and $\mathbf{C}$ connected both with internal
structure of the point, and with its interaction with external
fields. It is shown in \S 3 that equation of motion of free mass
point reduces to a conservation of the velocity of the center of
inertia. To obtain solutions natural equations of motion are
introduced for internal degrees of freedom describing precession
of pseudo-vectors $\mathbf{S}$ and $\mathbf{C}$ round the
direction of the velocity of the center of inertia. All solutions
of the equation of motion for a free mass point (when $\mathbf{S}
= \mathbf{S}_0$ and $\mathbf{C} = \mathbf{C}_0$) in the
center-of-inertia reference frame are found in \S 4. The center of
inertia proves to be does not coincide with the center of mass. As
a result, the center of mass moves by a complicated trajectory
round the direction of motion of the center of inertia. Some
solutions in the center-of-inertia reference frame are infinite.
It is shown that they become finite at zero energy of the mass
point. It is of interest that equation of motion admits also
solutions for zero mass and transversal polarization. In \S 5
equation of moments is considered and speculations are contained
about the physical sense of pseudo-vectors $\mathbf{S}_0$ and
$\mathbf{C}_0$ and their relation with spin whose equation of
motion in general case we deal with in \S 6, which contains also
conclusive remarks on possible interpretation of obtained
solutions.

\section{Description of the mass point with internal degrees of
freedom}

A mass point with internal degrees of freedom can be considered as
a non-inertial system whose equation of motion taking into account
its interaction with an external field has the form of the
Newton's Second Law (~\cite{Tar1}-\cite{Tar2})
$$
\frac{d\mathbf{P}}{dt} = \mathbf{F} \;, \eqno{(2.1)}
$$
where
$$
{\bf P} = m_{0}{\bf V} - \frac{\partial U}{\partial {\bf V}} +
[{\bf S}\times {\bf W}] \;, \eqno{(2.2)}
$$
is a dynamical momentum of the point, $m_0$ is its rest mass,
$$
{\bf F} = - \frac{\partial U}{\partial {\bf R}} + [{\bf C}\times
{\bf V}] \;, \eqno{(2.3)}
$$
is a force, acting to the point. Expressions (2.2) and (2.3)
follow from the definition of the elementary work of force, $dA =
(\mathbf{F} \cdot d\mathbf{R})$, if potential function $U$ depends
on velocity. Potential function characterizes both medium in which
the point moves and interaction of the point with physical objects
which are in this medium. Therefore it should be assumed that $U$
generally can depend on time $t$, relative coordinates $\bf R$,
velocity $\bf V$ and accelerations ${\bf W}^{(k)} = d^{k}{\bf W} /
dt^{k}$, $k = 0,1,2,...,N$, so that $U = U(t,{\bf R},{\bf V},{\bf
W},{\bf \dot {W}},...,{\bf W}^{(N)})$.

Internal degrees of freedom are characterized by pseudo-vectors
$\mathbf{S}$ and $\mathbf{C}$ connected with both internal
structure of mass point and interaction. Hence, they can be
represented as sums
$$
{\bf S} = {\bf S}_0 + {\bf S}^{ext} \;, \; {\bf C} = {\bf C}_0 +
{\bf C}^{ext} \;, \eqno{(2.4)}
$$
where ${\bf S}^{ext}$ and ${\bf C}^{ext}$ are connected
exclusively with interaction and depend on the same variables as
potential function; ${\bf S}_0$ and ${\bf C}_0$ are connected
exclusively with internal structure of mass point and during its
motion they can change only in the direction but not in the module
provided an interaction is neglected.

If the function $U$ is represented in the form
$$
U = U_0 - ({\bf R} \cdot [{\bf V}\times{\bf C}]) = U_0 + ({\bf V}
\cdot [{\bf R}\times{\bf C}]) = U_0 - ([{\bf R} \times {\bf V}]
\cdot {\bf C}) \;, \eqno{(2.5)}
$$
then Eqs. (2.2) and (2.3) take the form
$$
{\bf P} = m_{0}{\bf V} - \frac{\partial U_0}{\partial {\bf V}} +
[{\bf S}\times {\bf W}] - [{\bf R}\times {\bf C}] + ([\mathbf{R}
\times \mathbf{V}] \cdot \frac{\partial \mathbf{C}^{ext}}{\partial
{\bf V}}) \;, \eqno{(2.6)}
$$
$$
{\bf F} = - \frac{\partial U_0}{\partial {\bf R}} + ([\mathbf{R}
\times \mathbf{V}] \cdot \frac{\partial \mathbf{C}^{ext}}{\partial
{\bf R}}) \;, \eqno{(2.7)}
$$
where
$$
([{\bf R}\times {\bf V}] \cdot \frac{\partial{\bf
C}^{ext}}{\partial{\bf R}})_i = \varepsilon_{klm} R^{k} V^{l}
\frac{\partial({\bf C}^{ext})^m}{\partial{R^i}} \;, \eqno{(2.8)}
$$
$$
([{\bf R}\times {\bf V}] \cdot \frac{\partial{\bf
C}^{ext}}{\partial{\bf V}})_i = \varepsilon_{klm} R^{k} V^{l}
\frac{\partial({\bf C}^{ext})^m}{\partial{V^i}} \;. \eqno{(2.9)}
$$

It follows from Eq.(2.1), which is reduced to equation
$$
\frac{d}{dt}\left( {m_{0} {\bf V} + [{{\bf S}_0}\times {\bf W}] -
[{\bf R}\times{{\bf C}_0}]} \right) = - \frac{\partial
U_0}{\partial {\bf R}} + ([{\bf R}\times {\bf V}] \cdot
\frac{\partial{\bf C}^{ext}}{\partial{\bf R}}) +
$$
$$
 + \frac{d}{dt} \left( \frac{\partial U_0}{\partial
{\bf V}} - ([{\bf R}\times {\bf V}] \cdot \frac{\partial{\bf
C}^{ext}}{\partial{\bf V}}) - [{{\bf S}^{ext}}\times {\bf W}] +
[{\bf R}\times{{\bf C}^{ext}}] \right) \;, \eqno{(2.10)}
$$
that there take place an equation for energy
$$
 \frac{dE}{dt} = \frac{\partial U}{\partial t} + \sum\limits_{k =
 0}^N {(\frac{\partial U}{\partial {\bf W}^{(k)}} \cdot {\bf
 W}^{(k + 1)})} = \frac{\partial U_0}{\partial t} - ([\mathbf{R}
 \times \mathbf{V}] \cdot \frac{\partial \mathbf{C}^{ext}}{\partial
 t}) +
$$
$$
 + \sum\limits_{k = 0}^N {(\frac{\partial U_0}{\partial {\bf
 W}^{(k)}} \cdot {\bf W}^{(k + 1)})} - \sum\limits_{k = 0}^N
 {(([\mathbf{R} \times \mathbf{V}] \cdot \frac{\partial
 \mathbf{C}^{ext}}{\partial {\bf W}^{(k)}}) \cdot {\bf W}^{(k +
 1)})} \;,\eqno{(2.11)}
$$
where
$$
E = \frac{m_{0}{\bf V}^2}{2} - ([{\bf V} \times {\bf W}] \cdot
{\bf S}_0) - ([{\bf V} \times {\bf W}] \cdot {\bf S}^{ext}) -
({\bf V} \cdot \frac{\partial U_0}{\partial {\bf V}})+
$$
$$
 + ({\bf V} \cdot ([\mathbf{R} \times \mathbf{V}] \cdot
 \frac{\partial \mathbf{C}^{ext}}{\partial \mathbf{V}})) + U_0 \;
 , \eqno{(2.12)}
$$
$$
 ({\bf V} \cdot ([\mathbf{R} \times \mathbf{V}] \cdot
 \frac{\partial \mathbf{C}^{ext}}{\partial \mathbf{V}})) =
 \varepsilon_{klm} R^{k} V^{l} V^{i} \frac{\partial
 (\mathbf{C}^{ext})^{m}}{\partial V^{i}} \;, \eqno{(2.13)}
$$
$$
 (([\mathbf{R} \times \mathbf{V}] \cdot
 \frac{\partial \mathbf{C}^{ext}}{\partial \mathbf{W}^{(k)}})
 \cdot {\bf W}^{(k+1)}) = \varepsilon_{klm} R^{k} V^{l}
 (\mathbf{W}^{(k+1)})^{i} \frac{\partial
 (\mathbf{C}^{ext})^{m}}{\partial (\mathbf{W}^{(k)})^{i}} \;.
 \eqno{(2.14)}
$$
Obviously, the energy (2.12) is integral of motion, if r.h.s. of
Eq. (2.11) becomes zero.

\section{The motion of free mass point}

In this section we will consider a motion of free mass point M
with internal degrees of freedom, and will not concern its
interaction and physical sense of quantities $\mathbf{S}$ and
$\mathbf{C}$. Mass point in question will be free, if $U_0 = 0$,
${\bf S}^{ext} = 0$, ${\bf C}^{ext} = 0$. Then equation (2.10)
gives rise to conservation of the vector
$$
{\bf P}_{\mathrm{C}} = m_{0}{\bf V} + [{\bf S}_{0} \times {\bf W}]
- [{\bf R}\times {\bf C}_0] = m{\bf V}_{\mathrm{C}} \;,
\eqno{(3.1)}
$$
where $m_{0}{\bf V}$ is kinetic momentum of the point M being
\textit{a center of mass}, $m$ is an effective mass. It is
reasonably to term the vector (3.1) as the kinetic momentum
associated with the point M. Here $\textbf{V}_{\mathrm{C}} = d
\textbf{R}_{\mathrm{C}}/ dt$ is a velocity of some point C
specified by radius-vector
$$
{\bf R}_{\mathrm{C}}(t) = {\bf R}_{0} + {\bf V}_{\mathrm{C}}t \;,
\eqno{(3.2)}
$$
where ${\bf R}_{0}$ is a radius-vector of initial position of the
point C.

It follows from Eq.(2.7) that the vector $\mathbf{P}_{\mathrm{C}}$
is a constant vector, if $U_0$ does not depend on the relative
radius-vector. According to Eq.(3.2) the point C moves inertially
with velocity ${\bf V}_{\mathrm{C}}$. Hence, it is \textit{a
center of inertia}, which in general does not coincide with the
center of mass M, moving along some trajectory round the direction
of ${\bf V}_{\mathrm{C}}$. Figure 1 shows parameters of this
trajectory.
\begin{center}
\includegraphics[scale=0.7,keepaspectratio,draft=false]{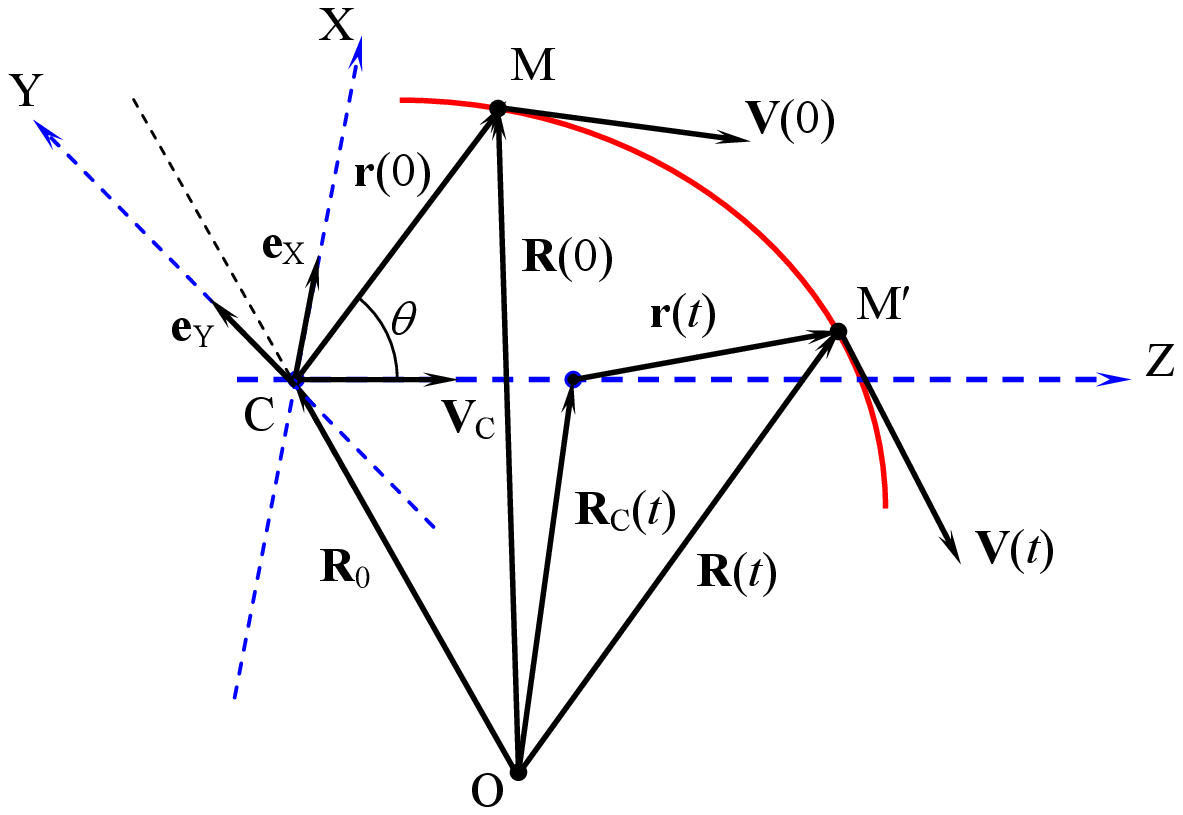}
\end{center}
\centerline{Figure 1. Parameters of the trajectory of the mass
point M} \vspace{5mm}

Eq.(3.1) may be rewritten in the form
$$
m_{0}{\bf v} + [{\bf S}_{0} \times {\bf w}] - [{\bf R}\times {\bf
C}_0] = (m - m_0){\bf V_{\mathrm{C}}}\;, \eqno{(3.3)}
$$
where
$$
{\bf r}(t) = {\bf R}(t) - {\bf R}_{\mathrm{C}}(t) \;, \; {\bf
v}(t) = {\bf V}(t) - {\bf V}_{\mathrm{C}} \;, \; {\bf w}(t) = {\bf
W}(t) \; \eqno{(3.4)}
$$
are radius-vector, velocity and acceleration of the center of mass
M relative to the center of inertia C, respectively.

The energy (2.12) can be expressed as
$$
E = K_{\mathrm{C}} + K_{0{\mathrm{C}}} + E_0 \;, \eqno{(3.5)}
$$
where
$$
K_{\mathrm{C}} = \frac{m_{0} {\bf V}^{2}_{\mathrm{C}}}{2} \;
\eqno{(3.6)}
$$
is a kinetic energy of the center of inertia as though the mass of
the point M was in the point C,
$$
K_{0{\mathrm{C}}} = m_{0} ({\bf v} \cdot {\bf V}_{\mathrm{C}}) +
({\bf V}_{\mathrm{C}} \cdot [\textbf{S}_{0} \times \textbf{w}]) \;
\eqno{(3.7)}
$$
is additional kinetic energy stipulated by both the motion of the
center of inertia and the motion of the point M relative to the
center of inertia C,
$$
E_{0} = \frac{m_{0} {\bf v}^{2}}{2} + (\textbf{v} \cdot
[\textbf{S}_{0} \times \textbf{w}]) \; \eqno{(3.8)}
$$
is a kinetic energy of the point M stipulated by its motion
relative to the center of inertia.

According to equation of motion (3.1) radius-vector
$\textbf{R}(t)$ should be determined by pseudo-vectors
$\textbf{S}_{0}$ and $\textbf{C}_{0}$, which are precessing with
the same velocity about the direction of the vector ${\bf
V}_{\mathrm{C}}$. Therefore $\textbf{S}_{0}$ and $\textbf{C}_{0}$
ought to satisfy to following equations of motion
$$
\frac{d \textbf{S}_0}{dt} = [\mathbf{\Omega}_{0} \times
\textbf{S}_0] \;, \frac{d \textbf{C}_0}{dt} = [\mathbf{\Omega}_{0}
\times \textbf{C}_{0}] \;, \eqno{(3.9)}
$$
where
$$
\mathbf{\Omega}_{0} = \sigma \mathbf{V}_\mathrm{C} = \Omega_{0}
\mathbf{e}_{Z} \;,  \eqno{(3.10)}
$$
is an angular velocity of precession, $\sigma = \mathrm{const}$
has a dimension of inverse length. If we choose the axis Z to be
coincided with the direction of ${\bf V}_{\mathrm{C}}$, i.e. ${\bf
V}_{\mathrm{C}} = V_{\mathrm{C}} \mathbf{e}_{Z}$, then $\Omega_{0}
= \sigma V_{\mathrm{C}}$ can be both positive, and negative
quantity. Unit vectors $\mathbf{e}_{X}$ and $\mathbf{e}_{Y}$ may
be chosen in the following form
$$
\mathbf{e}_{X} = K^{1/2} [\mathbf{r}_{0} \times
\mathbf{\Omega}_{0}] \;, \eqno{(3.11)}
$$
$$
\mathbf{e}_{Y} = \Omega^{-1}_{0} K^{1/2} [\mathbf{\Omega}_{0}
\times [\mathbf{r}_{0} \times \mathbf{\Omega}_{0}]] \;,
\eqno{(3.12)}
$$
where
$$
K = [\mathbf{\Omega}_{0} \times \mathbf{r}_0]^{-2} =
\frac{1}{\mathbf{r}_0^{2} \mathbf{\Omega}_{0}^{2} \sin^{2} \theta}
\;, \eqno{(3.13)}
$$
$$
[\mathbf{\Omega}_{0} \times [\mathbf{r}_0 \times
\mathbf{\Omega}_{0}]]^{2} = \mathbf{\Omega}_{0}^{2} K^{-1} \;,
\eqno{(3.14)}
$$
$\mathbf{r}_0 = \mathbf{r}(0)$ is a radius-vector of the center of
mass M relative to the center of inertia C at initial time $t =
0$; $\theta = \pi /2$ corresponds to that the vector
$\mathbf{r}_0$ lies in a plane, perpendicular to a direction of
motion.

Equations (3.9) have solutions
$$
\mathbf{S}_{0} = S_{0}( \sin{\alpha_S} \sin{\Omega_{0} t}
\mathbf{e}_{X} + \sin{\alpha_S} \cos{\Omega_{0} t} \mathbf{e}_{Y}
+ \cos{\alpha_S} \mathbf{e}_{Z}) \;, \eqno{(3.15)}
$$
$$
\mathbf{C}_{0} = C_{0}( \sin{\alpha_C} \sin{\Omega_{0} t}
\mathbf{e}_{X} + \sin{\alpha_C} \cos{\Omega_{0} t} \mathbf{e}_{Y}
+ \cos{\alpha_C} \mathbf{e}_{Z}) \;, \eqno{(3.16)}
$$
where $S_{0} = |\mathbf{S}_{0}| = \mathrm{const}$, $C_{0} =
|\mathbf{C}_{0}| = \mathrm{const}$, $\alpha_S$ and $\alpha_C$ are
constant angles between $\mathbf{S}_{0}$, $\mathbf{C}_{0}$ and
direction of $\mathbf{V}_{\mathrm{C}}$, respectively.

It is convenient to solve equations (3.3) in the center-of-inertia
reference frame, where $\mathbf{R}_{0} = \mathbf{0}$,
$\mathbf{R}_{\mathrm{C}} = \mathbf{0}$, $\mathbf{V}_{\mathrm{C}} =
\mathbf{0}$, $E = E_0$. Let us introduce dimensionless variable
$\xi = \Omega_{0} t$ and denotations
$$
\mu_{S} = 1 + \frac{2 \Omega_{0} S_{0}}{m_{0}} \cos{\alpha_S}
\;,\; \lambda_{S} = \frac{2 \Omega_{0} S_{0}}{m_{0}}
\sin{\alpha_S} \;,\; \eqno{(3.17)}
$$
$$
\mu_{C} = 1 - \frac{2 C_{0}}{m_{0} \Omega_{0}} \cos{\alpha_C}
\;,\; \lambda_{C} = \frac{2 C_{0}}{m_{0} \Omega_{0}}
\sin{\alpha_C} \;,\; \eqno{(3.18)}
$$
$$
\varepsilon_{0} = \frac{2 E_0}{m_{0} \mathbf{r}_{0}^{2}
\Omega_{0}^2} \;. \eqno{(3.19)}
$$

Then, representing $\mathbf{r}(t)$ in the form
$$
\mathbf{r}(\xi) = r_{0} [\sin{\xi} \mathbf{e}_{X} + \cos{\xi}
\mathbf{e}_{Y} + B(\xi)\mathbf{e}_{Z}] Z(\xi) \;, \eqno{(3.20)}
$$
where functions $B(\xi)$ and $Z(\xi)$ satisfy to initial condition
$$
B(0) = \cot{\theta} \;,\; Z(0) = \frac{|[\mathbf{r}_{0} \times
\mathbf{\Omega}_{0}]|}{r_{0} \Omega_{0}} = \sin{\theta} \;,
\eqno{(3.21)}
$$
we reduce equation of motion (3.3) and equation of energy (3.8) to
following system
$$
(\lambda_{S} B - \mu_{S}+1) Z'' + 2\lambda_{S} B'Z' + (\lambda_{S}
B'' + \lambda_{C} B + \mu_{S} + \mu_{C}) Z = 0 \;, \eqno{(3.22)}
$$
$$
\mu_{S} Z' = 0 \;, \eqno{(3.23)}
$$
$$
[(B - \lambda_{S}) Z]' = 0 \;. \eqno{(3.24)}
$$
$$
 (\lambda_{S} B - \mu_{S} + 1) ZZ'' + (B^{2} - 2 \lambda_{S} B + 2
 \mu_{S} - 1) Z'^{2} +
$$
$$ + 2 BB'ZZ' + (\lambda_{S} B'' + B'^{2} + \mu_{S}) Z^{2} =
\varepsilon_{0} \;. \eqno{(3.25)}
$$

\section{ Solutions of the equation of motion}

The detailed analysis of equations (3.22)-(3.25) leads to
following possible solutions (~\cite{Tar3}).

\textbf{I.} $m_{0} \neq 0$, $\mathbf{S}_{0} \neq \mathbf{0}$,
$\mathbf{C}_{0} \neq \mathbf{0}$.

\textbf{I.1.} $B=0$, $Z=1$, $\mu_{S} \neq 0$,
 $\lambda_{S} \neq 0$, $\lambda_{C} \neq 0$, $\mu_{C} = -\mu_{S} =
 -\varepsilon_{0}$.

$$
\cos{\alpha_{S}} = \frac{2 E_{0} - m_{0} r_{0}^{2}
\Omega_{0}^{2}}{2 r_{0}^{2} \Omega_{0}^{3} S_{0}} \;, \eqno{(4.1)}
$$
$$
\cos{\alpha_{C}} = \frac{2 E_{0} + m_{0} r_{0}^{2}
\Omega_{0}^{2}}{2 r_{0}^{2} \Omega_{0} C_{0}} \;. \eqno{(4.2)}
$$

The equation of a trajectory looks like
$$
\mathbf{r}(t) = r_{0} [\sin{\Omega_{0} t} \mathbf{e}_{X} +
\cos{\Omega_{0} t} \mathbf{e}_{Y}] \;, \eqno{(4.3)}
$$
i.e. the point M moves on a circle round Z-direction with angular
velocity $\Omega_{0}$. The direction of pseudo-vector
$\mathbf{S}_{0}$ is given by Eq.(4.1), and the direction of
pseudo-vector $\mathbf{C}_{0}$ is determined by Eq.(4.2),
following from the condition $\mu_{C} = -\mu_{S}$. An angle
$\beta$ between $\mathbf{S}_{0}$ and $\mathbf{C}_{0}$ may be found
from the relation
$$
\cos{\beta} = \frac{(\mathbf{S}_{0} \cdot \mathbf{C}_{0})}{S_{0}
C_{0}} = \frac{1}{S_{0} C_{0}} \left(\frac{E_{0}^{2}}{r_{0}^{4}
\Omega_{0}^{4}} - \frac{m_{0}^{2}}{4} \right) + \Biggl[1 -
\frac{1}{\Omega_{0}^{2} S_{0}^{2}}
\left(\frac{E_{0}^{2}}{r_{0}^{4} \Omega_{0}^{4}} - \frac{m_{0}
E_{0}}{r_{0}^{2} \Omega_{0}^{2}} + \frac{m_{0}^{2}}{4} \right) -
$$
$$
- \frac{\Omega_{0}^{2}}{C_{0}^{2}}
\left(\frac{E_{0}^{2}}{r_{0}^{4} \Omega_{0}^{4}} + \frac{m_{0}
E_{0}}{r_{0}^{2} \Omega_{0}^{2}} + \frac{m_{0}^{2}}{4} \right) +
\frac{1}{S_{0}^{2} C_{0}^{2}} \left(\frac{E_{0}^{2}}{r_{0}^{4}
\Omega_{0}^{4}} - \frac{m_{0}^{2}}{4} \right )^{2}
\Biggr]^{\frac{1}{2}} \;. \eqno{(4.4)}
$$

\textbf{I.2.} $B=0$, $Z=1$, $\mu_{S} = 0$,
 $\lambda_{S} \neq 0$, $\lambda_{C} \neq 0$, $\mu_{C} = 0$,
 $\varepsilon_{0} = 0$. This case is obtained from the previous one at
 $E_{0} = 0$.
$$
\cos{\alpha_{S}} = -\frac{m_{0}}{2 \Omega_{0} S_{0}} \;,
\eqno{(4.5)}
$$
$$
\cos{\alpha_{C}} = \frac{m_{0} \Omega_{0}}{2 C_{0}} \;.
\eqno{(4.6)}
$$

The trajectory is described by Eq.(4.3) and represents a circle of
radius $r_{0}$, lying in a plane, perpendicular to a direction of
the motion of the center of inertia C. A direction of the motion
of the point M in the cases I.1, I.2 is determined by sign of
angular velocity of precession $\Omega_{0}$, consistent with
equations (4.1) and (4.2). For example, for $E_{0} = 0$  and
$\cos{\alpha_{S}} > 0$ we have $\Omega_{0} < 0$, whence it follows
$\cos{\alpha_{C}} < 0$. Hence, such a point moves by left spiral
directed along $\mathbf{V}_{\mathrm{C}}$. An angle $\beta$ between
$\mathbf{S}_{0}$ and $\mathbf{C}_{0}$ may be found from the
relation

$$
\cos{\beta} = \left [1 - \frac{m_{0}^{2} (\Omega_{0}^{4} S_{0}^{2}
+ C_{0}^{2})}{4 \Omega_{0}^{2} S_{0}^{2} C_{0}^{2}} +
\frac{m_{0}^{4}}{16 S_{0}^{2} C_{0}^{2}} \right]^{\frac{1}{2}} -
\frac{m_{0}^{2}}{4 S_{0} C_{0}} \;. \eqno{(4.7)}
$$

\textbf{I.3.} $B=0$, $Z = \pm \xi \sqrt{-\varepsilon_{0}} + 1$,
$\mu_{S} = 0$,
 $\lambda_{S} = 0$, $\lambda_{C} \neq 0$, $\mu_{C} = 0$,
 $\varepsilon_{0} \leq 0$.
$$
\alpha_{S} = 0 \;, \; S_{0} = - \frac{m_{0}}{2 \Omega_{0}} \;, \;
\Omega_{0} < 0 \;, \; \; \; \mathrm{or} \; \; \; \; \alpha_{S} =
\pi \;,\, S_{0} = \frac{m_{0}}{2 \Omega_{0}} \;, \; \Omega_{0} > 0
\;; \eqno{(4.8)}
$$
$$
\cos{\alpha_{C}} = \frac{m_{0} \Omega_{0}}{2 C_{0}} \;,
\eqno{(4.9)}
$$
$$
\cos{\beta} = - \frac{m_{0}^{2}}{4 S_{0} C_{0}} \;. \eqno{(4.10)}
$$

The equation of a trajectory looks like
$$
\mathbf{r}(t) = r_{0} [\sin{\Omega_{0} t} \mathbf{e}_{X} +
\cos{\Omega_{0} t} \mathbf{e}_{Y}] (\pm \Omega_{\varepsilon} t +1)
\;, \eqno{(4.11)}
$$
where $\Omega_{\varepsilon} = \Omega_{0} \sqrt{-\varepsilon_{0}}$.
In the center-of-inertia reference frame the trajectory is a plane
helix, perpendicular to the direction of
$\mathbf{V}_{\mathrm{C}}$. In the laboratory reference frame the
trajectory is convergent and then divergent conical spiral. At
$E_{0} = 0$ the trajectory (4.11) becomes finite and takes the
form (4.3).

\textbf{I.4.} $B=0$, $\mu_{S} = 0$,
 $\lambda_{S} = 0$, $\lambda_{C} \neq 0$, $\mu_{C} = -\varepsilon_{0}$.
$$
\alpha_{S} = 0 \;, \; S_{0} = - \frac{m_{0}}{2 \Omega_{0}} \;, \;
\Omega_{0} < 0 \;, \; \; \; \mathrm{or} \; \; \; \; \alpha_{S} =
\pi \;, \; S_{0} = \frac{m_{0}}{2 \Omega_{0}} \;, \; \Omega_{0} >
0 \;; \eqno{(4.12)}
$$
$$
\cos{\alpha_{C}} = \frac{2 E_{0} + m_{0} r_{0}^{2}
\Omega_{0}^{2}}{2 C_{0} r_{0}^{2} \Omega_{0}} \;, \eqno{(4.13)}
$$
$$
\cos{\beta} = - \frac{2 m_{0} E_{0} + m_{0}^{2} r_{0}^{2}
\Omega_{0}^{2}}{4 S_{0} C_{0} r_{0}^{2} \Omega_{0}^{2}} \;;
\eqno{(4.14)}
$$
$$
Z(\xi) = \cos{\xi \sqrt{-\varepsilon_{0}}} \;, \; \;
\varepsilon_{0} < 0 \;, \eqno{(4.15)}
$$
$$
Z(\xi) = \cosh{\xi \sqrt{\varepsilon_{0}}} \;, \; \;
\varepsilon_{0} \geq 0 \;, \eqno{(4.16)}
$$

The equations of a trajectory are
$$
\mathbf{r}(t) = r_{0} [\sin{\Omega_{0} t} \mathbf{e}_{X} +
\cos{\Omega_{0} t} \mathbf{e}_{Y}] \cos{\Omega_{\varepsilon} t}
\;, \; E_{0} < 0 \;, \eqno{(4.17)}
$$
$$
\mathbf{r}(t) = r_{0} [\sin{\Omega_{0} t} \mathbf{e}_{X} +
\cos{\Omega_{0} t} \mathbf{e}_{Y}] \cosh{\Omega_{\varepsilon} t}
\;, \; E_{0} \geq 0 \;,\eqno{(4.18)}
$$
Here, as in the preceding case infinite trajectories (4.18) become
finite ones at $E_{0} = 0$.

\textbf{I.5.} $B=0$, $\mu_{S} = 0$,
 $\lambda_{S} = 0$, $\lambda_{C} = 0$, $\mu_{C} \neq 0$.
$$
\alpha_{S} = 0 \;,\; S_{0} = - \frac{m_{0}}{2 \Omega_{0}} \;, \;
\Omega_{0} < 0 \;, \; \; \; \mathrm{or} \; \; \; \; \alpha_{S} =
\pi \;,\; S_{0} = \frac{m_{0}}{2 \Omega_{0}} \;, \; \Omega_{0} > 0
\;; \eqno{(4.19)}
$$
$$
\alpha_{C} = 0 \;,\; \mu_{C} = 1 - \frac{2 C_{0}}{m_{0}
\Omega_{0}} \;, \; \; \; \mathrm{or} \; \; \; \; \alpha_{C} = \pi
\;,\, \mu_{C} = 1 + \frac{2 C_{0}}{m_{0} \Omega_{0}} \;;
\eqno{(4.20)}
$$
$$
\cos{\beta} = \left\{%
\begin{array}{ll}
    \; \; \; \cos{\alpha_{C}} \;, \; \Omega_{0} < 0 \;, & \hbox{} \\
    - \cos{\alpha_{C}} \;, \; \Omega_{0} > 0 \;; & \hbox{} \\
\end{array}%
\right. \; \eqno{(4.21)}
$$
$$
Z(\xi) = \cos{(\xi \sqrt{\mu_{C}})} \pm
\sqrt{\frac{-\varepsilon_{0}}{\mu_{C}} - 1} \sin{(\xi
\sqrt{\mu_{C}})} \;, \; 0 < \mu_{C} < -\varepsilon_{0} \;, \;
\varepsilon_{0} < 0 \;, \eqno{(4.22)}
$$
$$
Z(\xi) = \cosh{(\xi \sqrt{-\mu_{C}})} \pm
\sqrt{\frac{-\varepsilon_{0}}{\mu_{C}} - 1} \sinh{(\xi
\sqrt{-\mu_{C}})} \;, \; \mu_{C} \leq 0 \;, \; \mu_{C} \leq
-\varepsilon_{0} \;. \eqno{(4.23)}
$$

The equations of a trajectory are
$$
\mathbf{r}(t) = r_{0} \sqrt{\frac{-\varepsilon_{0}}{\mu_{C}}} \;
[\sin{\Omega_{0} t} \mathbf{e}_{X} + \cos{\Omega_{0} t}
\mathbf{e}_{Y}] \cos{(\omega_{C} t \mp \varphi_{C})} \;, \; 0 <
\mu_{C} < -\varepsilon_{0} \;, \; E_{0} < 0 \;, \eqno{(4.24)}
$$
$$
\mathbf{r}(t) = r_{0} \sqrt{2 - \frac{-\varepsilon_{0}}{\mu_{C}}}
\; [\sin{\Omega_{0} t} \mathbf{e}_{X} + \cos{\Omega_{0} t}
\mathbf{e}_{Y}] \cosh{(\omega_{C} t \pm \varphi_{C})} \;, \;
\mu_{C} \leq 0 \;, \; \mu_{C} \leq -\varepsilon_{0} \;,
\eqno{(4.25)}
$$
where
$$
\omega_{C} = \sqrt{\mu_{C}} \Omega_{0} \;, \; \varphi_{C} =
\arctan{\sqrt{\frac{-\varepsilon_{0}}{\mu_{C}} - 1}} \;, \; 0 <
\mu_{C} < -\varepsilon_{0} \;, \eqno{(4.26)}
$$
$$
\omega_{C} = \sqrt{-\mu_{C}} \Omega_{0} \;, \; \varphi_{C} =
\tanh^{-1}{\sqrt{\frac{-\varepsilon_{0}}{\mu_{C}} - 1}} \;, \;
\mu_{C} \leq 0 \;, \; \mu_{C} \leq -\varepsilon_{0} \;,
\eqno{(4.27)}
$$
Infinite trajectories (4.25) can be excluded, having imposed a
condition $\mu_{C} = \varepsilon_{0} = 0$. Then (4.25) reduces to
(4.3).

\textbf{I.6.} $B Z = \lambda_{S} (Z - 1)$ \;,
$\mu_{S} = 0$ \;, $\lambda_{S} = - \tan{\alpha_{S}} =
\sqrt{\frac{4 \Omega_{0}^{2} S_{0}^{2}}{m_{0}^2} - 1}$ \;,
$\lambda_{C} = 0$ \;,

\parindent=50pt\ $\mu_{C} = 0$ \;, $\varepsilon_{0} < 0$ \;.
$$
\cos{\alpha_{S}} = - \frac{m_{0}}{2 \Omega_{0} S_{0}} \;,
\eqno{(4.28)}
$$
$$
\alpha_{C} = 0 \;,\; C_{0} = \frac{m_{0} \Omega_{0}}{2} \;, \;
\Omega_{0} > 0 \;, \; \; \; \mathrm{or} \; \; \; \; \alpha_{C} =
\pi \;,\, C_{0} = -\frac{m_{0} \Omega_{0}}{2} \;, \; \Omega_{0} <
0 \;; \eqno{(4.29)}
$$
$$
\cos{\beta} = - \frac{m_{0}}{2 \Omega_{0} S_{0}} \cos{\alpha_{C}}
\;, \eqno{(4.30)}
$$
$$
Z(\xi) = \pm \xi \sqrt{\frac{-\varepsilon_{0}}{\lambda_{S}^{2}
+1}} + 1 \;. \eqno{(4.31)}
$$

The equation of a trajectory is
$$
\mathbf{r}(t) = r_{0} [\sin{\Omega_{0} t} \mathbf{e}_{X} +
\cos{\Omega_{0} t} \mathbf{e}_{Y}] \Biggl[\pm
\frac{\sqrt{-2m_{0}E_{0}}}{2r_{0} \Omega_{0}^{2} S_{0}} \Omega_{0}
t + 1 \Biggr] \pm
$$
$$
 \pm \sqrt{-\frac{2 E_{0}}{m_{0}\Omega_{0}^{2}}
(1 - \frac{m_{0}^{2}}{4 \Omega_{0}^{2} S_{0}^{2}})} \Omega_{0} t
\mathbf{e}_{Z} \;, \eqno{(4.32)}
$$
reducing to finite form (4.3) at $E_{0} = 0$.

\textbf{I.7.} $B Z = \lambda_{S} (Z - 1)$ \;, $\mu_{S} = 0$ \;,
$\lambda_{S} = - \tan{\alpha_{S}} = \sqrt{\frac{4 \Omega_{0}^{2}
S_{0}^{2}}{m_{0}^2} - 1}$ \;, $\lambda_{C} = 0$ \;, $\mu_{C} \neq
0$ \;.
$$
\cos{\alpha_{S}} = - \frac{m_{0}}{2 \Omega_{0} S_{0}} \;,
\eqno{(4.33)}
$$
$$
\alpha_{C} = 0 \;,\; \mu_{C} = 1 - \frac{2 C_{0}}{m_{0}
\Omega_{0}} \;, \; \; \; \mathrm{or} \; \; \; \; \alpha_{C} = \pi
\;,\, \mu_{C} = 1 + \frac{2 C_{0}}{m_{0} \Omega_{0}} \;;
\eqno{(4.34)}
$$
$$
\cos{\beta} = - \frac{m_{0}}{2 \Omega_{0} S_{0}} \cos{\alpha_{C}}
\;; \eqno{(4.35)}
$$
$$
Z(\xi) = \cos{\sqrt{\frac{\mu_{C}}{\lambda_{S}^{2} +1}} \xi} \pm
\sqrt{\frac{-\varepsilon_{0}}{\mu_{C}} -1}
\sin{\sqrt{\frac{\mu_{C}}{\lambda_{S}^{2} +1}} \xi} \;, \; 0 <
\mu_{C} < -\varepsilon_{0} \;, \; \varepsilon_{0} < 0 \;,
\eqno{(4.36)}
$$
$$
Z(\xi) = \cosh{\sqrt{\frac{-\mu_{C}}{\lambda_{S}^{2} +1}} \xi} \pm
\sqrt{\frac{-\varepsilon_{0}}{\mu_{C}} -1}
\sinh{\sqrt{\frac{-\mu_{C}}{\lambda_{S}^{2} +1}} \xi} \;, \;
\mu_{C} \leq 0 \;, \; \mu_{C} \leq -\varepsilon_{0} \;.
\eqno{(4.37)}
$$

The equations of a trajectory are
$$
\mathbf{r}(t) = r_{0} \sqrt{\frac{-\varepsilon_{0}}{\mu_{C}}} \;
[\sin{\Omega_{0} t} \mathbf{e}_{X} + \cos{\Omega_{0} t}
\mathbf{e}_{Y}] \cos{(\Omega_{C} t \pm \varphi_{C})} +
$$
$$ + r_{0} \sqrt{\frac{4 \Omega_{0}^{2}
S_{0}^{2}}{m_{0}^{2}} -1} \left[
\sqrt{\frac{-\varepsilon_{0}}{\mu_{C}}} \cos{(\Omega_{C} t \pm
\varphi_{C})} -1 \right] \mathbf{e}_{Z} \;, \; 0 < \mu_{C} <
-\varepsilon_{0} \;; \eqno{(4.38)}
$$
$$
\mathbf{r}(t) = r_{0} \sqrt{2 - \frac{-\varepsilon_{0}}{\mu_{C}}}
\; [\sin{\Omega_{0} t} \mathbf{e}_{X} + \cos{\Omega_{0} t}
\mathbf{e}_{Y}] \cosh{(\Omega_{C} t \pm \varphi_{C})} +
$$
$$
+ r_{0} \sqrt{\frac{4 \Omega_{0}^{2} S_{0}^{2}}{m_{0}^{2}} -1}
\left[ \sqrt{2 - \frac{-\varepsilon_{0}}{\mu_{C}}} \;
\cosh{(\Omega_{C} t \pm \varphi_{C})} -1 \right] \mathbf{e}_{Z}
\;, \; \mu_{C} \leq 0 \;, \; \mu_{C} \leq -\varepsilon_{0} \;,
\eqno{(4.39)}
$$
where $\varphi_{C}$ is determined in Eqs.(4.26)-(4.27),
$$
\Omega_{C} = \omega_{C} \cos{\alpha_{S}} \;. \eqno{(4.40)}
$$
Infinite trajectories (4.39) becomes finite ones, when $\mu_{C} =
\varepsilon_{0} = 0$. Then the equation (4.39) takes the form
$$
\mathbf{r}(t) = r_{0} [\sin{\Omega_{0} t} \mathbf{e}_{X} +
\cos{\Omega_{0} t} \mathbf{e}_{Y}] - r_{0} \sqrt{\frac{4
\Omega_{0}^{2} S_{0}^{2}}{m_{0}^{2}} -1} \mathbf{e}_{Z} \;, \;
\eqno{(4.41)}
$$
and it follows from (4.34) that
$$
\alpha_{C} = 0 \;,\; C_{0} = \frac{m_{0} \Omega_{0}}{2} \;, \;
\Omega_{0} >0 \;, \; \; \mathrm{or} \; \; \; \; \alpha_{C} = \pi
\;,\, C_{0} = - \frac{m_{0} \Omega_{0}}{2} \;, \; \Omega_{0} < 0
\;. \eqno{(4.42)}
$$

\textbf{II.} $m_{0} \neq 0$, $\mathbf{S}_{0} \neq \mathbf{0}$,
$\mathbf{C}_{0} = \mathbf{0}$.

In this case equation (2.10) is a generalization of
non-relativistic Frenkel-Mathisson-Weyssenhoff equation
(~\cite{Fre},~\cite{Mat2},~\cite{Wey}), describing the motion of
point particle with constant spin $\mathbf{s} = -c^{2}
\mathbf{S}_{0}$. It may be deduced from the cases I.5 and I.7 at
$\mu_{C} = 1$, $\lambda_{C} = 0$. As a result we have following
variants.

\parindent=24pt\ \textbf{II.1.} $B = 0$ \;, $\mu_{S} = 0$ \;,
$\lambda_{S} = 0$ \;, $\lambda_{C} = 0$ \;, $\mu_{C} = 1$ \;.
$$
\alpha_{S} = 0 \;,\; S_{0} = - \frac{m_{0}}{2 \Omega_{0}} \;, \;
\Omega_{0} < 0 \;, \; \; \; \mathrm{or} \; \; \; \; \alpha_{S} =
\pi \;,\; S_{0} = \frac{m_{0}}{2 \Omega_{0}} \;, \; \Omega_{0} > 0
\;; \eqno{(4.43)}
$$
$$
Z(\xi) = \cos{\xi} \pm \sqrt{-\varepsilon_{0}-1} \sin{\xi} \;, \;
\varepsilon_{0} < 0 \;. \eqno{(4.44)}
$$
The equation of a trajectory looks as
$$
\mathbf{r}(t) = r_{0} \sqrt{-\varepsilon_{0}} \; [\sin{\Omega_{0}
t} \mathbf{e}_{X} + \cos{\Omega_{0} t} \mathbf{e}_{Y}]
\cos{(\Omega_{0} t \mp \varphi)} \;, \; E_{0} < 0 \;,
\eqno{(4.45)}
$$
where
$$
\varphi = \arctan{\sqrt{-\varepsilon_{0} -1}} \;. \eqno{(4.46)}
$$

\parindent=24pt\ \textbf{II.2.} $B Z = \lambda_{S} (Z - 1)$, \;
$\mu_{S} = 0$, \; $\lambda_{S} = - \tan{\alpha_{S}} =
\sqrt{\frac{4 \Omega_{0}^{2} S_{0}^{2}}{m_{0}^{2}} - 1}$, \;
$\lambda_{C} = 0$, \; $\mu_{C} = 1$.
$$
\cos{\alpha_{S}} = - \frac{m_{0}}{2 \Omega_{0} S_{0}} \;,
\eqno{(4.47)}
$$
$$
Z(\xi) = \cos{\frac{\xi}{\sqrt{\lambda_{S}^{2} +1}}} \pm
\sqrt{-\varepsilon_{0} -1} \sin{\frac{\xi}{\sqrt{\lambda_{S}^{2}
+1}}} \;, \; \varepsilon_{0} < 0 \;. \eqno{(4.48)}
$$

\parindent=24pt\ The equation of a trajectory is
$$
\mathbf{r}(t) = r_{0} \sqrt{-\varepsilon_{0}} \; [\sin{\Omega_{0}
t} \mathbf{e}_{X} + \cos{\Omega_{0} t} \mathbf{e}_{Y}]
\cos{(\Omega_{0} t \cos{\alpha_{S} \mp \varphi)}} +
$$
$$
+ r_{0} \sqrt{\frac{4 \Omega_{0}^{2} S_{0}^{2}}{m_{0}^{2}} -1}
[\sqrt{-\varepsilon_{0}} \cos{(\Omega_{0} t \cos{\alpha_{S} \mp
\varphi)}} - 1] \mathbf{e}_{Z} \;, \; E_{0} < 0 \;. \eqno{(4.49)}
$$

\parindent=24pt\ The case II.1 is deduced from the case II.2 at
$\lambda_{S} = 0$, i.e. if conditions (4.43) fulfill. Assuming for
the electron $S_{0} = s/c^{2} = \hbar / 2 c^{2}$, $m_{0} = m_{e}$,
we obtain $h \nu_{Z} = \hbar \Omega_{Z} = m_{e} c^{2}$, or
$$
\nu_{Z} = \frac{\Omega_{Z}}{2 \pi} = \frac{|\Omega_{0}|}{2 \pi}
\approx 1.236 \cdot 10^{20} \; \mathrm{Hz} \eqno{(4.50)}
$$
is the frequency corresponding to generally accepted rest energy
of the electron or its Compton wavelength $\lambda_{e} = h / m_{e}
c = 2.42627 \cdot 10^{-12} \; \mathrm{m}$. The energy of the
electron in the center-of-inertia reference frame is negative,
$E_{0} = - m_{e} r_{0}^{2} \Omega_{0}^{2} / 2$.

Polarization, or spin projection to the direction of motion, is
determined by the value
$$
P = \frac{(\mathbf{S}_{0} \cdot \mathbf{V}_{\mathrm{C}})}{S_{0}
V_{\mathrm{C}}} = \frac{(\mathbf{S}_{0} \cdot
\mathbf{\Omega})}{\sigma S_{0} V_{\mathrm{C}}} = \cos{\alpha_{S}}
\;. \eqno{(4.51)}
$$
Value $P = +1$ corresponds to $\alpha_{S} = 0$, $\Omega_{0} < 0$,
i.e. to counter-clockwise motion, whereas $P = -1$ corresponds to
$\alpha_{S} = \pi$, $\Omega_{0} > 0$, i.e. to clockwise motion. It
suggests associating these motions with motions of the electron
and positron, which hence should differ from each other by the
type of motion rather than by charge. More strictly it it is
possible to prove or refuse this hypothesis, having considered
two-body problem taking into account their interaction with each
other, as well as their motion in constant electric and magnetic
fields.

\textbf{III.} $m_{0} \neq 0$, $\mathbf{S}_{0} = \mathbf{0}$,
$\mathbf{C}_{0} \neq \mathbf{0}$. In this case $\lambda_{S} = 0$,
$\mu_{S} = 1$, and equations (3.22)-(3.25) give solutions $B = 0$,
$Z = 1$, and relations $\varepsilon_{0} = 1$, $\mu_{C} = -1$, or
$$
E_{0} = \frac{m_{0} r_{0}^{2} \Omega_{0}^{2}}{2} \;, \;
\cos{\alpha_{C} = \frac{m_{0} \Omega_{0}}{C_{0}}} \;.
\eqno{(4.52)}
$$

It is easy to see that this case turns out from case I.1 at
$\lambda_{S} = 0$. The equation of a trajectory looks like
Eq.(4.3).

\textbf{IV.} The system of equations (3.22)-(3.25) admits the
solution corresponding to zero mass, $m_{0} = 0$. In this case
$\mu_{C} = 1$, $\lambda_{C} = 0$, $Z = 1$, and equation of motion
gives following variants.

\textbf{IV.1.} $m_{0} = 0$, $\mathbf{S}_{0} \neq \mathbf{0}$,
$\mathbf{C}_{0} \neq \mathbf{0}$.
$$
B = 0 \;, \; \cos{\alpha_{S} = \frac{E_{0}}{r_{0}^{2}
\Omega_{0}^{3} S_{0}}} \;, \; \cos{\alpha_{C} =
\frac{E_{0}}{r_{0}^{2} \Omega_{0} C_{0}}}\;. \eqno{(4.53)}
$$
In the center-of-inertia the equation of a trajectory looks like
Eq.(4.3).

\textbf{IV.2.} $m_{0} = 0$, $\mathbf{S}_{0} \neq \mathbf{0}$,
$\mathbf{C}_{0} = \mathbf{0}$.
$$
B(\xi) = -\frac{\cot{\alpha_{S}}}{2} \xi^{2} + B_{1} \xi \;, \;
B_{1} = \mathrm{const} \;, \; E_{0} = 0 \;. \eqno{(4.54)}
$$
In the center-of-inertia the equation of a trajectory looks like
$$
\mathbf{r}(t) = r_{0} [\sin{\Omega_{0} t} \mathbf{e}_{X} +
\cos{\Omega_{0} t} \mathbf{e}_{Y}] + r_{0}
\left[-\frac{\cot{\alpha_{S}}}{2} \Omega_{0}^{2} t^{2} + B_{1}
\Omega_{0} t \right] \mathbf{e}_{Z} \;. \eqno{(4.55)}
$$

To eliminate such divergent trajectories, it is sufficient to
assume $B_{1} = 0$, $\alpha_{S} = \pm \pi /2$, what corresponds to
transversal polarization of pseudo-vector $\mathbf{S}_{0}$.

\textbf{IV.3.} $m_{0} = 0$, $\mathbf{S}_{0} = \mathbf{0}$,
$\mathbf{C}_{0} \neq \mathbf{0}$.

In this case we have $B = 0$, $\alpha_{C} = \pm \pi /2$, $E_{0} =
0$, what corresponds to transversal polarization of pseudo-vector
$\mathbf{C}_{0}$. The equation of a trajectory looks like
Eq.(4.3).

Summarizing the results obtained, it is possible to assert that
all finite trajectories of free mass points with internal degrees
of freedom are subdivided into three types.

Trajectories of the first type are right or left helix along the
direction of motion of the center of inertia. They are specific
for the cases I.1-I.3, I.6, I.7 (at $E_{0} = 0$), III and IV, and
there is no restrictions in $E_0$ only for the cases I.1 and IV.1,
whereas for the rest cases we have $E_{0} = 0$. Figure 2 shows
trajectories for the case I.3 for polarization $P = -1$ (clockwise
motion along Z-axis, Figure 2a) and $P = +1$ (counter-clockwise
motion along Z-axis, Figure 2b).

Trajectories of the second type in the center-of-inertia reference
frame are plane multi-petal rosettes. They are specific for the
cases I.4, I.5, I.7 (at $S_{0}^{2} = m_{0}^{2} / 4
\Omega_{0}^{2}$) and II.1. They are closed $2N$-petal plane
rosettes, when frequencies $\Omega_{\varepsilon}$, $\omega_{C}$
and $\Omega_{C}$ are multiple to the frequency $\Omega_{0}$ and
represented in Figure 3 for the case I.4 ($E_{0} \leq 0$,
$\Omega_{\varepsilon} = 2 \Omega_{0}$). As in preceding cases the
motion is clockwise for $P = -1$ (Figure 3a) and counter-clockwise
one for $P = +1$ (Figure3b). The direction of pseudo-vector
$\mathbf{C}_{0}$ in Figures 2 and 3 does not pointed out.

\begin{center}
\includegraphics[scale=0.9,keepaspectratio,draft=false]{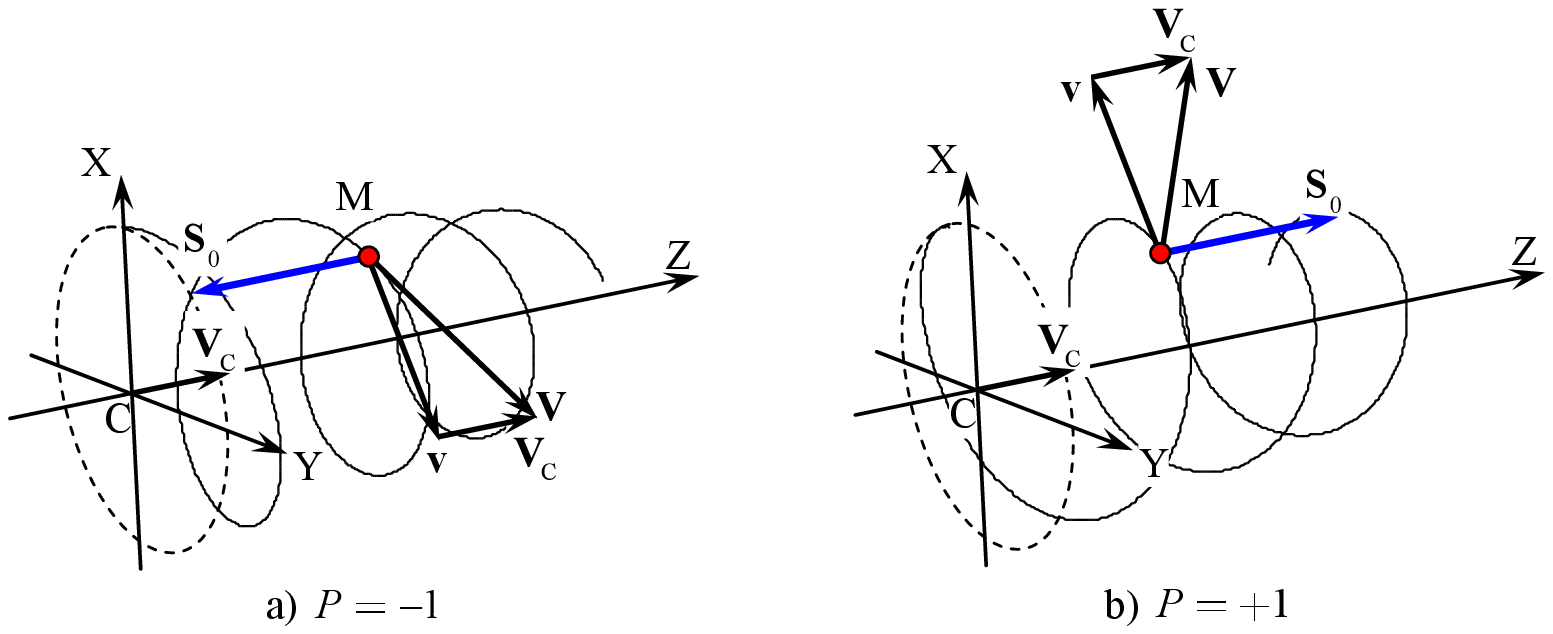}
\end{center}
\vspace{-20mm} \centerline{Figure 2.}
\begin{center}
\includegraphics[scale=0.9,keepaspectratio,draft=false]{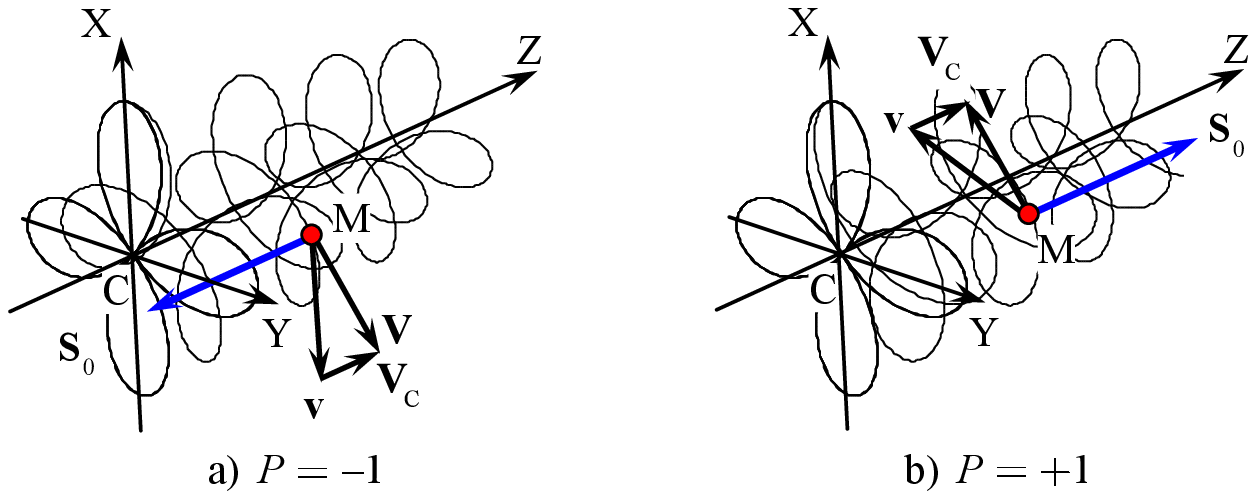}
\end{center}
\vspace{-10mm} \centerline{Figure 3.}
\begin{center}
\includegraphics[scale=0.9,keepaspectratio,draft=false]{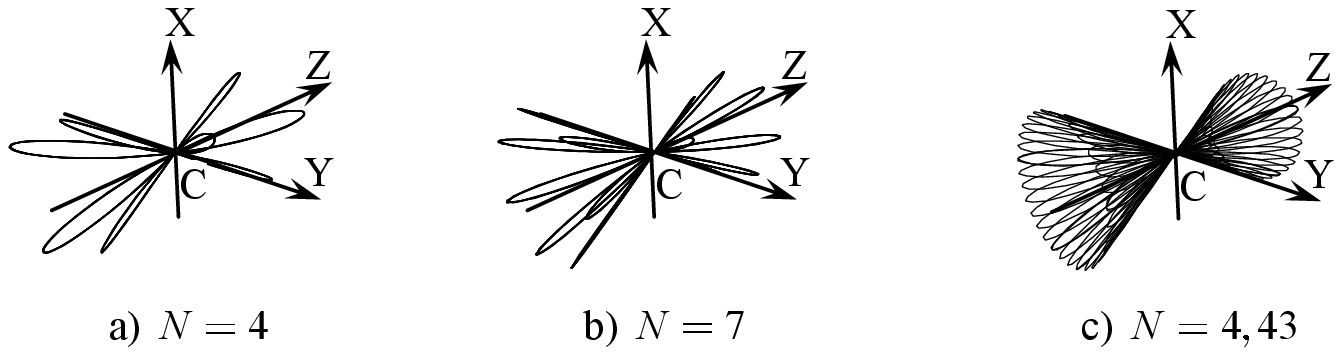}
\end{center}
\vspace{-15mm} \centerline{Figure 4.} \vspace{5mm}

Trajectories of the third type are specific for the cases I.7 and
II.2. They are multi-petal symmetric space curves in the
center-of-inertia reference frame. The number of petals is defined
by a condition of the maximum removal of the mass point from the
center of inertia that takes place at $\cos{(\Omega_{C} t \pm
\varphi_{C})} = 1$, i.e. at
$$
\frac{m_{0} \sqrt{\mu_{C}}}{2 S_{0}} t \pm
\arctan{\sqrt{\frac{-\varepsilon_{0}}{\mu_{C}} - 1}} = 2 k \pi \;,
\; \; k = 0, \pm 1, \pm 2, ... \;. \eqno{(4.56)}
$$
These curves are closed $2N$-petal space rosettes, when frequency
$\Omega_{C}$ is multiple to the frequency $\Omega_{0}$,
$\Omega_{C} = N \Omega_{0}$, or
$$
\frac{m_{0} \sqrt{\mu_{C}}}{2 S_{0} \Omega_{0}} = \sqrt{\mu_{C}}
\cos{\alpha_{S}} = N \;. \eqno{(4.57)}
$$
The motion is going clockwise for $P < 0$ and counter-clockwise
for $P < 0$. Figure 4 shows samples of trajectories for $P < 0$,
$N = 4$ (Figure 4a), $N = 7$ (Figure 4b) and for the nonintegral
$N = 4.43$ (Figure 4c) in the interval $0 \leq \Omega_{0} t \leq
50$.

\section{ Equation of moments and definition of spin}

As it is known, one of internal property of particles is spin,
associated classically with proper angular momentum of particle.
Therefore a temptation arises to connect pseudo-vectors
$\mathbf{S}$ and $\mathbf{C}$ with spin. For the sake of it we
will consider the equation of moments
$$
\frac{d \mathbf{L}}{dt} = \mathbf{M} + \mathbf{T}, \eqno{(5.1)}
$$
where
$$
\mathbf{L} \doteq [\mathbf{R} \times \mathbf{P}] = m_{0}
[\mathbf{R} \times \mathbf{V}] - [\mathbf{R} \times \frac{\partial
U}{\partial \mathbf{V}}] + [\mathbf{R} \times [\mathbf{S} \times
\mathbf{W}]] =
$$
$$
= m_{0} [\mathbf{R} \times \mathbf{V}] - [\mathbf{R} \times
\frac{\partial U_0}{\partial \mathbf{V}}] + [\mathbf{R} \times
[\mathbf{S} \times \mathbf{W}]] - [\mathbf{R} \times [\mathbf{R}
\times \mathbf{C}]] + [\mathbf{R} \times ([\mathbf{R} \times
\mathbf{V}] \cdot \frac{\partial \mathbf{C}^{ext}}{\partial
\mathbf{V}})] \eqno{(5.2)}
$$
is a dynamical angular momentum,
$$
\mathbf{M} \doteq [\mathbf{R} \times \mathbf{F}] = - [\mathbf{R}
\times \frac{\partial U}{\partial \mathbf{R}}] + [\mathbf{R}
\times [\mathbf{C} \times \mathbf{V}]] =
$$
$$
= - [\mathbf{R} \times \frac{\partial U_0}{\partial \mathbf{R}}] +
[\mathbf{R} \times ([\mathbf{R} \times \mathbf{V}] \cdot
\frac{\partial \mathbf{C}^{ext}}{\partial \mathbf{R}})]
\eqno{(5.3)}
$$
is a moment of force, acting at the mass point,
$$
\mathbf{T} \doteq [\mathbf{V} \times \mathbf{P}] = - [\mathbf{V}
\times \frac{\partial U}{\partial \mathbf{V}}] + [\mathbf{V}
\times [\mathbf{S} \times \mathbf{W}]] =
$$
$$
= - [\mathbf{V} \times \frac{\partial U_0}{\partial \mathbf{V}}] +
[\mathbf{V} \times [\mathbf{S} \times \mathbf{W}]] - [\mathbf{V}
\times [\mathbf{R} \times \mathbf{C}]] + [\mathbf{V} \times
([\mathbf{R} \times \mathbf{V}] \cdot \frac{\partial
\mathbf{C}^{ext}}{\partial \mathbf{V}})] \eqno{(5.4)}
$$
is an additional twisting moment, or torque.~\footnote{In standard
mechanics the concept "torque" is applied sometimes to the moment
of force (5.3).  Here we distinguish the moment of force (5.3) and
torque (5.4).}

The equation of moments (5.1) follows from equation of motion
(2.1). Therefore, having only definition (5.2) for angular
momentum it is impossible to define a concept of proper angular
momentum for the point with internal degrees of freedom. Indeed,
let us consider free mass point, for which $U_{0} = 0$,
$\mathbf{S}^{ext} = \mathbf{0}$, $\mathbf{C}^{ext} = \mathbf{0}$.
Then the moment of force (5.3) becomes zero, and the dynamical
angular momentum (5.2) and torque (5.4) are equal to
$$
\mathbf{L} = m_{0} [\mathbf{R} \times \mathbf{V}] + [\mathbf{R}
\times [\mathbf{S}_{0} \times \mathbf{W}]] - [\mathbf{R} \times
[\mathbf{R} \times \mathbf{C}_{0}]] \;, \eqno{(5.5)}
$$
$$
\mathbf{T} = [\mathbf{V} \times [\mathbf{S}_{0} \times
\mathbf{W}]] - [\mathbf{V} \times [\mathbf{R} \times
\mathbf{C}_{0}]] \;. \eqno{(5.6)}
$$
respectively.

Introducing variables relative to the center of inertia (3.4), we
have
$$
\mathbf{L} = \mathbf{L}_{\mathrm{C}} + \mathbf{L}_{0 \mathrm{C}} +
\mathbf{L}_{0} \;, \eqno{(5.7)}
$$
$$
\mathbf{T} = \mathbf{T}_{\mathrm{C}} + \mathbf{T}_{0 \mathrm{C}} +
\mathbf{T}_{0} \;, \eqno{(5.8)}
$$
where
$$
\mathbf{L}_{\mathrm{C}} = m_{0} [\mathbf{R}_{\mathrm{C}} \times
\mathbf{V}_{\mathrm{C}}] - [\mathbf{R}_{\mathrm{C}} \times
[\mathbf{R}_{\mathrm{C}} \times \mathbf{C}_{0}]] \eqno{(5.9)}
$$
is an angular momentum of the center of inertia C relative to the
origin O, as if total rest-mass $m_0$ was in the center of inertia
C,
$$
\mathbf{L}_{0 \mathrm{C}} = m_{0} [\mathbf{r} \times
\mathbf{V}_{\mathrm{C}}] + m_{0} [\mathbf{R}_{\mathrm{C}} \times
\mathbf{v}] + [\mathbf{R}_{\mathrm{C}} \times [\mathbf{S}_{0}
\times \mathbf{w}]] - [\mathbf{R}_{\mathrm{C}} \times [\mathbf{r}
\times \mathbf{C}_{0}]] - [\mathbf{r} \times
[\mathbf{R}_{\mathrm{C}} \times \mathbf{C}_{0}]] \eqno{(5.10)}
$$
is an angular momentum of the center of mass M relative to the
origin O, related with both its motion relative to the center of
inertia C and a motion of the latter one in absolute reference
frame,
$$
\mathbf{L}_{0} = m_{0} [\mathbf{r} \times \mathbf{v}] +
[\mathbf{r} \times [\mathbf{S}_{0} \times \mathbf{w}]] -
[\mathbf{r} \times [\mathbf{r} \times \mathbf{C}_{0}]]
\eqno{(5.11)}
$$
is an angular momentum of the center of mass M relative to the
center of inertia C;
$$
\mathbf{T}_{\mathrm{C}} = - [\mathbf{V}_{\mathrm{C}} \times
[\mathbf{R}_{\mathrm{C}} \times \mathbf{C}_{0}]] \eqno{(5.12)}
$$
is a torque relative to the origin O, acting upon the center of
mass M,
$$
\mathbf{T}_{0 \mathrm{C}} = [\mathbf{V}_{\mathrm{C}} \times
[\mathbf{S}_{0} \times \mathbf{w}]] - [\mathbf{V}_{\mathrm{C}}
\times [\mathbf{r} \times \mathbf{C}_{0}]] - [\mathbf{v} \times
[\mathbf{R}_{\mathrm{C}} \times \mathbf{C}_{0}]] \eqno{(5.13)}
$$
is additional torque relative to the origin O, acting upon the
center of mass M and related with both its motion relative to the
center of inertia C and a motion of the latter one in absolute
reference frame,
$$
\mathbf{T}_{0} = [\mathbf{v} \times [\mathbf{S}_{0} \times
\mathbf{w}]] - [\mathbf{v} \times [\mathbf{r} \times
\mathbf{C}_{0}]] \eqno{(5.14)}
$$
is a torque relative to the center of inertia C, acting upon the
center of mass M.

In the center-of-inertia reference frame we have
$\mathbf{L}_{\mathrm{C}} = \mathbf{0}$, $\mathbf{L}_{0 \mathrm{C}}
= \mathbf{0}$, $\mathbf{T}_{\mathrm{C}} = \mathbf{0}$,
$\mathbf{T}_{0 \mathrm{C}} = \mathbf{0}$, so that the equation of
moments (5.1) takes the form
$$
\frac{d \mathbf{L}_{0}}{dt} = \mathbf{T}_{0} \;, \eqno{(5.15)}
$$
or
$$
[\mathbf{r} \times \frac{d}{dt} [m_{0} \mathbf{v} +
[\mathbf{S}_{0} \times \mathbf{w}] -[\mathbf{r} \times
\mathbf{C}_{0}]]] = \mathbf{0} \;. \eqno{(5.16)}
$$
In the reference frame in question $m_{0} \mathbf{v} +
[\mathbf{S}_{0} \times \mathbf{w}] -[\mathbf{r} \times
\mathbf{C}_{0}] = \mathbf{0}$, therefore equations (5.15) and
hence (5.1) are identities.

On the other hand, since $\mathbf{r} = \mathbf{0}$, $\mathbf{v} =
\mathbf{0}$, $\mathbf{w} = \mathbf{0}$, in the center-of-mass
reference frame, then $\mathbf{L}_{0} = \mathbf{0}$. Therefore
$\mathbf{L}_{0}$ cannot play a role of proper angular momentum
(spin) of the point M to which we want to relate pseudo-vectors
$\mathbf{S}$ and $\mathbf{C}$. For definition of their physical
sense additional reasons are necessary. Point M with internal
degrees of freedom should be considered as non-inertial extended
object rotating with angular velocity $\boldsymbol\omega_{0}$ and
possessing the proper angular momentum (spin). To take into
account internal rotational degrees of freedom it is necessary to
introduce the total moment of momentum instead of angular moment
(5.7) (see, for example,~\cite{Cor})
$$
\mathbf{J} = \mathbf{L} + \mathbf{s} = \mathbf{L}_{\mathrm{C}} +
\mathbf{L}_{0 \mathrm{C}} + \mathbf{L}_{0} + \mathbf{s} \;,
\eqno{(5.17)}
$$
which is defined as spin $\mathbf{s}$ in the center-of-mass
reference frame, and equals to
$$
\mathbf{J}_{0} = \mathbf{L}_{0} + \mathbf{s} \eqno{(5.18)}
$$
in the center-of-inertia reference frame.

When interaction is missing, equation of motion (2.1) reduces to
Eq.(3.1). Consequently, pseudo-vectors $\mathbf{J}_{0}$,
$\mathbf{L}_{0}$ and $\mathbf{s}$ in the center-of-inertia
reference frame should be precessing round the direction of the
center of inertia with the same angular velocity
$\mathbf{\Omega}_{0} = \sigma \mathbf{V}_{\mathrm{C}}$, as
pseudo-vectors $\mathbf{S}_{0}$ and $\mathbf{C}_{0}$, i.e. they
have to satisfy to equations of motion of the same form
$$
\frac{d \mathbf{J}_{0}}{dt} = [\mathbf{\Omega}_{0} \times
\mathbf{J}_{0}] = \sigma [\mathbf{V}_{\mathrm{C}} \times
\mathbf{J}_{0}] \;, \eqno{(5.19)}
$$
$$
\frac{d \mathbf{L}_{0}}{dt} = [\mathbf{\Omega}_{0} \times
\mathbf{L}_{0}] = \sigma [\mathbf{V}_{\mathrm{C}} \times
\mathbf{L}_{0}] \;, \eqno{(5.20)}
$$
$$
\frac{d \mathbf{s}}{dt} = [\mathbf{\Omega}_{0} \times \mathbf{s}]
= \sigma [\mathbf{V}_{\mathrm{C}} \times \mathbf{s}] \;.
\eqno{(5.21)}
$$

Comparison of Eq.(5.20) with Eq.(5.15) and taking into account
Eq.(5.14) gives
$$
\frac{d \mathbf{L}_0}{dt} = [\mathbf{\Omega}_{0} \times
\mathbf{L}_0] = \mathbf{T}_0 = [\mathbf{v} \times ([\mathbf{S}_{0}
\times \mathbf{w}] - [\mathbf{r} \times \mathbf{C}_0])] = -
[\mathbf{v} \times m \mathbf{v}] = \mathbf{0} \;, \eqno{(5.22)}
$$
whence it follows that $\mathbf{L}_0$ is parallel to the angular
velocity of precession $\mathbf{\Omega}_0$,
$$
\mathbf{L}_{0} = m_{0} [\mathbf{r} \times \mathbf{v}] +
[\mathbf{r} \times [\mathbf{S}_{0} \times \mathbf{w}]] -
[\mathbf{r} \times [\mathbf{r} \times \mathbf{C}_{0}]] = I_{0}
\mathbf{\Omega}_{0} = m_{0} r_{0}^{2} \mathbf{\Omega}_{0} \;,
\eqno{(5.23)}
$$
whereas spin $\mathbf{s}$ is parallel to the angular velocity
$\boldsymbol\omega_0$ of proper rotation of extended point M, if
it is defined as its proper angular momentum. Then pseudo-vector
$\mathbf{L}_{0}$ gets a sense of orbital angular momentum of the
point M relative to the center of inertia. If $\mathbf{\hat{j}}_0$
is a proper tensor of inertia of the point M, the spin is defined
as follows
$$
\mathbf{s} = \mathbf{\hat{j}}_{0} \boldsymbol{\omega}_{0} = j_{0}
\boldsymbol{\omega}_{0} \;, \eqno{(5.24)}
$$
where $j_0$ is eigenvalue of $\mathbf{\hat{j}}_{0}$, i.e. proper
moment of inertia of the point M relative to the axis of its
rotation. Due to Eq.(5.23) and Eq.(5.24) the total moment of
momentum (5.18) is equal to
$$
\mathbf{J}_{0} = \mathbf{L}_{0} + \mathbf{s} = m_{0} r_{0}^{2}
\mathbf{\Omega}_{0} + j_{0} \boldsymbol{\omega}_{0} =
\mathbf{\hat{I}} \mathbf{\Omega}_{0} \;, \eqno{(5.25)}
$$
where $\mathbf{\hat{I}}$ is the tensor of inertia of the point M
relative to the center of inertia C, $r_0$ is the radius of
Zitterbewegung.

Applying the expression (5.24) to electron roughly represented as
a rigid sphere of radius $\rho_0$, consisting of structureless
mass points, we have $j_{0} = 2 m_{e} \rho_{0}^{2}/5$, $s =
\hbar/2$, whence it follows
$$
\omega_{0} = 2 \pi \nu_{0} = \frac{s}{j_0} = \frac{5 \hbar}{4
m_{e} \rho_{0}^{2}} \;. \eqno{(5.26)}
$$
Substituting here a value of the electron radius $\rho_{0} \approx
10^{-22} \; \mathrm{m}$ (Dehmelt,~\cite{Deh}), we will obtain an
estimate
$$
\nu_{0} = \frac{5 \hbar}{8 \pi m_{e} \rho_{0}^{2}} \approx 2.3
\cdot 10^{39} \; \mathrm{Hz} \;, \eqno{(5.27)}
$$
i.e. frequency of proper rotation of the electron is at least 19
orders greater than the frequency of Zitterbewegung (4.50),
$\nu_{Z} = m_{e} c^{2} /h \approx 1.24 \cdot 10^{20} \;
\mathrm{Hz}$, whereas a velocity on the electron surface is $v =
\nu_{0}\rho_{0} \approx 2.3 \cdot 10^{17} \;
\mathrm{m}/\mathrm{s}$, what is 9 orders greater than the speed of
light.

At high frequency of proper rotation in electron volume there
should be arising huge centrifugal forces of inertia, which
relocate interior substance of the electron to periphery. On the
other hand, stability of the electron implies that centrifugal
forces of inertia should be balanced by interior forces so that
the equilibrium shape of the electron represented something like a
toroid or a ring with a rigid surface. It is consistent with both
earlier idea by Parson-Compton~\cite{Com4},~\cite{Par}, and modern
toroidal or ring model (see, e.g.,~\cite{Ber},~\cite{Mat3}) or
dumbbell model of the electron (~\cite{Gri}). Ring (or dumbbell)
is characterized, at least, by two sizes, by its radius (length)
and thickness. Therefore it is difficult to say what value
$\rho_{0} \approx 10^{-22} \; \mathrm{m}$ obtained by Dehmelt
concerns. Even if $\rho_{0}$ is a classical electron radius,
$r_{e} = \alpha \lambda_{\mathrm{C}}/2 \pi \approx 2.9 \cdot
10^{-15} \; \mathrm{m}$, we will obtain instead of (5.27) an
estimate for the frequency of rotation $\nu_{0} \approx 2.7 \cdot
10^{24} \; \mathrm{Hz}$ and for the velocity $v = \nu_{0} r_{e}
\approx 7.8 \cdot 10^{9} \; \mathrm{m}/\mathrm{s}$, that also is
greater than the speed of light. In due time Lorentz has refused
an idea of extended electron as equatorial velocity of a surface
of spinning electron has turned out to be more than the speed of
light. However as long as we are in the frameworks of classical
mechanics, we have not any restriction on speed.

Whatever the electron would be actually arranged, its internal
structure should determine both field, created by it, and a type
of its motion, depending on a spin being integral property of this
structure. If the motion of free electron reduces to equations
(3.3), (3.9), then in the center-of-inertia reference frame we
obtain trajectories, described in \S 4. A motion of free extended
electron relative to the center of inertia means that, on the one
hand, its interior substance is acted upon by centrifugal forces
of inertia, and, on the other hand, by centripetal forces which
twist a trajectory. Resultant of these forces can be represented
as $m' [\mathbf{v} \times \mathbf{\Omega}_0]$, where $m'$ is some
coefficient with dimension of mass. For the point mass particle
there takes place a kind of equivalence principle, $m' = m_0$,
where the rest mass $m_0$ play a role of the measure of inertness
of inertially moving mass point, and a mass $m'$ is a measure of
non-inertiality of such a point moving with acceleration, whereas
a mass $m$, entering to right hand side of Eq.(3.1), is a measure
of inertness of extended object, which center of inertia moves
inertially. Therefore the hypothesis $m' = m_0$ is not obvious for
such extended particle as the electron, the more so mass of
micro-object depends on its interaction with external fields and
is not an additive quantity. Inasmuch as $m' [\mathbf{v} \times
\mathbf{\Omega}_0]$ is resultant force acting upon free electron,
it follows from equation (2.3), written in the center-of-inertia
reference frame,
$$
\mathbf{F} = \left(-\frac{\partial U}{\partial \mathbf{R}} +
[\mathbf{C} \times \mathbf{V}] \right)_{U = 0,
\mathbf{V}_{\mathrm{C}} = \mathbf{0}} = [\mathbf{C}_{0} \times
\mathbf{v}] = m' [\mathbf{v} \times \mathbf{\Omega}_0] \;.
\eqno{(5.28)}
$$
From here pseudo-vector $\mathbf{C}_{0}$ considering Eq.(3.10) may
be determined as
$$
\mathbf{C}_{0} = -m' \mathbf{\Omega}_{0} + \gamma \mathbf{v} = -m'
\sigma \mathbf{V}_{\mathrm{C}} + \gamma \mathbf{v} \;,
\eqno{(5.29)}
$$
where $\sigma$ and $\gamma$ are constant pseudo-scalars.

It should be noted that transformation (3.4) is a special case of
the Galileo transformation
$$
{\bf R'}(t) = {\bf R}(t) - {\bf V}_{\mathrm{K'}} t \;, \; {\bf
V'}(t) = {\bf V}(t) - {\bf V}_{\mathrm{K'}} \;, \; {\bf W'}(t) =
{\bf W}(t) \;. \eqno{(5.30)}
$$
where $\mathbf{V}_{\mathrm{K'}}$ is a velocity of inertial system
$\mathrm{K}'$ relative to absolute system K (here system
$\mathrm{K}'$ is the center-of-inertia reference frame, i.e.
$\mathbf{V}_{\mathrm{K'}} = \mathbf{V}_{\mathrm{C}}$). If the
Galileo's relativity principle is valid, equation (2.1) should be
covariant relative to transformations (5.30), i.e. in the system
$\mathrm{K}'$ it should be $d \mathbf{P'} / dt = \mathbf{F'}$,
where
$$
\mathbf{P} = m_{0} \mathbf{V} - \frac{\partial U}{\partial
\mathbf{V}} + [\mathbf{S} \times \mathbf{W}] = m_{0} \mathbf{V'} +
m_{0} \mathbf{V}_{\mathrm{K}'} - \frac{\partial U}{\partial
\mathbf{V'}} + [\mathbf{S} \times \mathbf{W'}] = \mathbf{P'} +
m_{0} \mathbf{V}_{\mathrm{K}'} \;, \eqno{(5.31)}
$$
$$
\mathbf{F} = - \frac{\partial U}{\partial \mathbf{R}} +
[\mathbf{C} \times \mathbf{V}] = - \frac{\partial U}{\partial
\mathbf{R'}} + [\mathbf{C} \times \mathbf{V'}] + [\mathbf{C}
\times \mathbf{V}_{\mathrm{K'}}] = \mathbf{F'} + [\mathbf{C}
\times \mathbf{V}_{\mathrm{K}'}] \;, \eqno{(5.32)}
$$
whence it follows relation
$$
[\mathbf{C} \times \mathbf{V}_{\mathrm{K}'}] = \mathbf{0} \;.
\eqno{(5.33)}
$$
For free electron we have $\mathbf{C} = \mathbf{C}_0$ and it
follows from Eq.(5.33) and Eq.(5.29) that $\gamma = 0$. Thus,
finally
$$
\mathbf{C}_{0} = -m' \mathbf{\Omega}_{0} = -m' \Omega_{0}
\mathbf{e}_{Z} \;. \eqno{(5.34)}
$$

Writing down expressions for $\mathbf{C}_{0}$, corresponding to
cases I.1- I.7 ($\mathbf{S}_{0} \neq \mathbf{0}$), and comparing
them with Eq.(5.34), we find that the condition (4.6),
corresponding to $\mu_{C} = 0$ and $E_{0} = 0$, should satisfied,
whence it follows $m' = - m_{0}/2$, and due to Eqs.(3.15), (4.1),
(4.5), (4.8), (4.12), (4.19), (4.28) and (4.33) pseudo-vector
$\mathbf{S}_{0}$ looks like
$$
\mathbf{S}_{0} = -\frac{m_0}{2 \Omega_0} \mathbf{e}_{Z} \;,
\eqno{(5.35)}
$$
i.e. two kinds of motion, corresponding to polarizations $P = \pm
1$, are possible. Combining Eqs.(5.34) and (5.35), we obtain
relation
$$
\mathbf{C}_{0} = -\Omega^{2}_{0} \mathbf{S}_{0} \;. \eqno{(5.36)}
$$

The case II ($\mathbf{C}_{0} = \mathbf{0}$, $E_{0} = -m_{0}
\mathbf{r}^{2}_{0} \Omega^{2}_{0}/2$) corresponds to $m' = 0$.

Expression for $\mathbf{C}_{0}$, corresponding to the case III
($\mathbf{S}_{0} = \mathbf{0}$, $E_{0} = m_{0} \mathbf{r}^{2}_{0}
\Omega^{2}_{0}/2$), which is similar to Eq.(5.34), gives $m' =
-m_0$. It may be assumed that the state with $\mathbf{S}_{0} =
\mathbf{0}$ is a bound state of two particles with opposite
polarizations, contribution of every of which in $\mathbf{C}_{0}$
is $m' = -m_{0}/2$. More strictly it can be confirmed after a
detailed solution of the two-body problem for interacting mass
points with internal degrees of freedom.

Trajectory of mass point in cases I and III is a circle (4.3) of
radius $r_0$ in the center-of-inertia reference frame. The
negative value of mass $m'$ means, that the force (5.28) is
centripetal, rather than centrifugal one. The unique reason of
such strange behavior is existence of internal rotational degrees
of freedom, described by pseudo-vector $\mathbf{S}_0$. It is
reasonably to express $\mathbf{S}_0$ in terms of spin (5.24), as
follows
$$
\mathbf{S}_{0} = - \frac{1}{c^2} \mathbf{s} \;, \eqno{(5.37)}
$$
where $c$ is some constant with dimension of velocity.

We obtain from Eqs.(5.24), (5.35) and (5.37)
$$
\mathbf{s} = j_{0} \boldsymbol{\omega}_{0} = \frac{m_{0} c^2}{2
\Omega_0} \mathbf{e}_{Z} = \frac{m_{0} c^2}{2 \Omega^{2}_{0}}
\mathbf{\Omega}_{0} \;. \eqno{(5.38)}
$$
Equations of motion (3.9) give $\mathbf{\dot{s}} = \mathbf{0}$,
implying conservation of spin direction when interaction is
negligible. Substituting (5.38) in (5.25), we obtain for the total
moment of momentum relative to the center of inertia
$$
\mathbf{J}_{0} = \left( 1+ \frac{2 r^{2}_{0} \Omega^{2}_0}{c^2}
\right) \mathbf{s} \;. \eqno{(5.39)}
$$
Introducing a denotation
$$
\hbar = \frac{m_{0} c^2}{|\Omega_0|} \;, \eqno{(5.40)}
$$
we have for spin $s = \hbar/2$. Here a question remains open
whether $c$ and $\hbar$ be the velocity of light and Planck
constant, respectively. Its solution will be determined by
behavior of particles in external fields and their interaction
with each other.

\section{Equation of motion for spin}

For obtaining complete solution of a problem about a motion of the
system in question it is necessary to add the equations for
internal degrees of freedom. In the previous paragraph we have
found out that the equation of the moments (5.1) is a consequence
of the equation (2.1), and it cannot be considered as the
additional equation. For free system we have the equation for spin
(5.21) which means that there exists a preferred direction,
namely, a direction of motion of the center of inertia, round
which the spin is precessing with constant angular velocity
$\mathbf{\Omega}_0$.

Generally at every given instant spin is precessing round any
instantaneous direction, simultaneously moving in space together
with the center of mass. If $\mathbf{N}(t)$ is a vector pointing
out in this direction the spin equation of motion can be written
as
$$
\frac{d \mathbf{s}}{dt} = [\mathbf{\Omega}_{\mathbf{N}} \times
\mathbf{s}] + \mathbf{m}(t) = \sigma_{\mathbf{N}}(t) [\mathbf{N}
\times \mathbf{s}] + \mathbf{m}(t) \;, \eqno{(6.1)}
$$
where $\mathbf{m}(t)$ is some pseudo-vector having a sense of the
moment of force or torque acting to extended point. The structure
of $\mathbf{m}(t)$, apparently, can be determined on specifying of
an interaction of internal substance of the point with external
fields. It follows from Eq.(6.1) the constancy of absolute value
of spin if $(\mathbf{m} \cdot \mathbf{s}) = 0$. Otherwise spin
changes not only over the direction, but also over absolute value.
Assuming the interaction of internal substance with external
fields to be much weaker than the interaction of the point as a
whole object we will consider $\mathbf{m} = \mathbf{0}$ as first
approximation. Besides, taking into account that equation (6.1)
should be reduced to Eq.(5.21), when interaction is negligible, it
is necessary to take as a vector $\mathbf{N}(t)$ the vector
$$
\mathbf{P} = m \mathbf{V}_{\mathrm{C}} = m_{0} \mathbf{V} -
\frac{\partial U_0}{\partial \mathbf{V}} - \frac{1}{c^2}
[\mathbf{s} \times (\mathbf{W} - \Omega^{2}_{0} \mathbf{R})] +
$$
$$
+ [\mathbf{S}^{ext} \times \mathbf{W}] - [\mathbf{R} \times
\mathbf{C}^{ext}] + ([\mathbf{R} \times \mathbf{V}] \cdot
\frac{\partial \mathbf{C}^{ext}}{\partial \mathbf{V}}) \;,
\eqno{(6.2)}
$$
which can be treated as a definition of the kinetic momentum
related to a particle, specifying the motion of the center of
inertia of the latter. In the presence of interaction the momentum
(6.2) is not conserved, and spin is precessing round instantaneous
direction of the momentum with instant angular velocity
$\mathbf{\Omega}(t) = \sigma_{\mathbf{P}}(t) \mathbf{P}/m$. Thus,
the equation of motion of the spin can be finally written down in
the form
$$
\frac{d \mathbf{s}}{dt} = \frac{\sigma_{\mathbf{P}}(t)}{m}
[\mathbf{P} \times \mathbf{s}] + \mathbf{m}(t) =
$$
$$
= \frac{m_{0} \sigma_{\mathbf{P}}(t)}{m} [\mathbf{V} \times
\mathbf{s}] - \frac{\sigma_{\mathbf{P}}(t)}{m} [\frac{\partial
U_0}{\partial \mathbf{V}} \times \mathbf{s}] +
\frac{\sigma_{\mathbf{P}}(t)}{m c^2} [\mathbf{s} \times
[\mathbf{s} \times (\mathbf{W} - \Omega^{2}_{0} \mathbf{R})]] -
$$
$$
- \frac{\sigma_{\mathbf{P}}(t)}{m} [\mathbf{s} \times
[\mathbf{S}^{ext} \times \mathbf{W}]] +
\frac{\sigma_{\mathbf{P}}(t)}{m} [\mathbf{s} \times [\mathbf{R}
\times \mathbf{C}^{ext}]] - \frac{\sigma_{\mathbf{P}}(t)}{m}
[\mathbf{s} \times ([\mathbf{R} \times \mathbf{V}] \cdot
\frac{\partial \mathbf{C}^{ext}}{\partial \mathbf{V}})] +
\mathbf{m}(t) \;. \eqno{(6.3)}
$$

All known equations of motion of spin in non-relativistic
approximation have structure of Eq.(6.3) with $\mathbf{m}(t) = 0$.
System of equations (2.1) and (6.3) with momentum (6.2) and force
(2.7) allow to solve a lot of problems about the motion of
spinning particles in various fields and to compare these
solutions with well-known results. Obtained here non-relativistic
equations of motion suppose the relativistic
generalization~\cite{Tar2}, however there are subtleties which
should be considered carefully.

Being founded on the stated above it is possible to assert that
the trembling motion of objects with internal degrees of freedom
has origin in classical mechanics. This circumstance does not
contradict numerous modern researches in which the classical
theory of spin is developing (see, e.g.,~\cite{Riv}). Experimental
observation of the motion of individual micro-objects with
internal structure is for now impossible at present status of
experimental technique. Nevertheless, the quantum phenomena of
Zitterbewegung type, which though are not experimentally
observable now, but they can be simulated, and the first results
are presented in work~\cite{Ger}. However, the results obtained
say that this phenomenon should take place, in principle, for such
classical objects as spinning top whose spin (proper moment of
momentum) is not parallel to the velocity of translational motion
of its center of inertia. Other examples are the projectile
(bullet), which is shot through a rifle trunk, and a boomerang. It
is possible to hope also that equations similar to the equations
obtained in \S 4 will allow explaining such little-understandable
phenomenon, as Dzhanibekov's effect. For micro-objects
Zitterbewegung is manifested so as they have the wave nature,
whereas this phenomenon is imperceptible for macro-objects because
of both their large mass, and smallness of radius (amplitude)
$r_{0}$ of trembling motion. Thus, there appears a possibility of
classical interpretation of quantum phenomena.

Moreover, two types of motion, corresponding to opposite
polarizations, give rise to new look at the origin of electric
charge. The solution of non-relativistic problem of the motion of
the point with internal degrees of freedom, given in \S\S 4-5, and
its application to free electron leads to new interpretation of
the charge of elementary particle which sign is determined by its
helicity. Right helicity $h = -P = +1$, $\alpha_{S} = \pi$,
$\Omega_{0} > 0$ (Figure 2a), corresponds to right polarization of
spin for free antiparticles (positrons), charged positively,
whereas left helicity $h = -P = -1$, $\alpha_{S} = 0$, $\Omega_{0}
< 0$ (Figure 2b), corresponds to left polarization of spin for
free particles (electrons), charged negatively. For interacting
particles helicity can be distinct from $h = \pm 1$, but its sign
as before corresponds to the sign of charge. Hence, \textit{the
charge is the conventional concept characterizing type of the
motion of spinning particle, corresponding to its helicity}. In
this connection it would like to mention W. Ritz's opinion,
according to which "these latter (electric charges) only playing,
like the masses in Mechanics, the role of coefficients,
conveniently chosen and invariable for a given ion or electron. In
a certain sense it is a mechanical theory of electricity"
(~\cite{Ritz}, p. 149).

\end{document}